\documentclass[a4paper]{JHEP3}

\usepackage{amsmath}
\usepackage{amssymb}

\title{Self-Protection of Massive Cosmological Gravitons}
\author{
Felix Berkhahn$^{abd}$, Dennis D.~Dietrich$^{c}$, and Stefan Hofmann$^{bd}$\\
\hskip -1.4mm$^a$Physik-Department T30d, Technische Universit\"at M\"unchen,
James-Franck-Stra{\ss}e, 85748 Garching, Germany\\
\hskip -1.4mm$^b$Excellence Cluster Universe, Boltzmannstra{\ss}e 2, 85748 Garching, Germany\\
\hskip -1.4mm$^c$CP$^\mathit{3}$-Origins, Centre for Particle Physics Phenomenology, University of Southern Denmark, Campusvej 55, 5230 Odense M, Denmark\\
\hskip -1.4mm$^d$Arnold Sommerfeld Center for Theoretical Physics, Ludwig-Maximilians-Universit\"at, Theresienstra{\ss}e 37, 80333 Munich, Germany\\~\\
E-Mails: \email{felix.berkhahn@ph.tum.de}, \email{dietrich@cp3.sdu.dk}, \email{stefan.hofmann@physik.lmu.de}
}
\abstract{
Relevant deformations of gravity present an exciting window of opportunity to 
probe the rigidity of gravity on cosmological scales. 
For a single-graviton theory, the leading relevant deformation constitutes a graviton mass term.
In this paper, we investigate the classical and quantum stability of massive cosmological
gravitons on generic Friedman backgrounds.  For a Universe expanding towards a de
Sitter epoch, we find that massive cosmological gravitons are self-protected against
unitarity violations by a strong coupling phenomenon. 
}%abstract

\keywords{gravity, modified gravity, quantum field theory on curved space, cosmological perturbation theory, dark energy}

\preprint{CP3-ORIGINS-2010-30\\LMU-ASC 61/10\\TUM-HEP-766/10}

\begin{document}

\maketitle

%%%%%%%%%%%%%%%%%%%%%%%%%%%%%%%%%%%%%%%%%%%%%%%%%%%%%%%%%%%%%%%%%%%%%%%%%%

\section{Introduction}

Technical naturalness is arguably one of the most promising pathfinders to physics 
beyond the standard model of particle interactions and gravity, as well.
It offers many exciting windows of opportunity related to renormalisable 
standard model operators that share an enhanced sensitivity to the scale
of new physics. 

Among these, the vacuum energy density is standing out in various ways. 
Being the unique operator with quartic sensitivity to the ultraviolet scale,
it also represents the most relevant term in the Einstein--Hilbert action. 
The basic observation is that the vacuum energy density is technically
unnatural within the standard model already at energy scales set by 
the lightest measured particle masses within its spectrum. 
In other words, the technical naturalness facet of this challenge is not solely
tied to the quantum gravity scale, unless there is an ultraviolet-infrared 
conspiracy operative in the vacuum sector that also respects the many
high-precision successes of the standard model at lower energies. 

Although collider experiments cannot measure the vacuum energy density, 
the challenge it poses becomes serious once the standard model of particle physics is coupled 
to gravity. Since gravity, being the most democratic field theory, 
couples to energy-momentum in a universal manner irrespective of its
sources' nature, the vacuum curves spacetime and affects the Universe's 
expansion history. 
As a consequence, and opening up yet another window of opportunity,
it seems attractive to reconsider gravity in the deep infrared and 
its consistent deformations.  

The hunt for a fundamental completion of gravity is characterised 
by incorporating new degrees of freedom in the ultraviolet with the 
infrared region kept untouched (in a relevant sense), its role reduced to
providing the classical benchmark tests. 
However, this precludes the opening up of additional gravitational degrees
of freedom that could be relevant in the deep infrared  on scales that are only 
poorly constrained by state-of-the-art cosmological observations.

Historically, this question was discouraged by a mighty no-go theorem 
stating the impossibility to embed gravity in a QCD-like theory with a 
self-interacting graviton multiplet under the spell of Yang--Mills. 
More precisely, assuming locality, Poincar\'e invariance, and a free-field 
limit consisting of massless gravitons, the only consistent deformations involving
a multiplet of gravitons are such that the deformed gauge algebra
is just a direct sum of independent diffeomorphism algebras \cite{Boulanger:2000bp}.

Interestingly, by relaxing one of its conditions --- allowing for new relevant 
degrees of freedom --- this no-go theorem gave rise to
a potent tool for studying consistent deformations of Einstein's gravity within
the effective field theory framework. For a nonzero deformation parameter,
the candidate deformations are characterised by a tremendous reduction
of symmetries:
\begin{equation}
\nonumber
{\rm diff}(M_1) \; \otimes \; {\rm diff}(M_2) \; \otimes \cdots
\longrightarrow
{\rm diag} \left({\rm diff}(M_1) \; \otimes \; {\rm diff}(M_2) \; \otimes \cdots\right)
\; ,
\end{equation}
with $M_j$ denoting (not necessarily) different spacetime (sub-)manifolds. 
The resulting gauge symmetry restricts the deformation, i.e.~the coupling
of different geometries, to be solely constructed from invariants of the reduced
symmetry  \cite{Damour:2002ws}. 

Under the umbrella of this framework, many proposals for relevant deformations 
of gravity that have been suggested in the last decade become cousins. 
Even more promising, the unique ghost-free theory for a massive spin-two field 
propagating all degrees of freedom \cite{Fierz:1939ix} fits under this umbrella. In the latter case,
two copies of Minkowski spacetime are considered, one perfect the other perturbed,
and the most relevant deformation becomes a spin-two mass term
\begin{equation}
\label{mdef}
S_{_{\rm deform}}
=
- \frac{m^2}{2} \int {\rm d}^4 x \; 
h_{\alpha\beta}\;  \mathcal{M}^{\alpha\beta\mu\nu} \; h_{\mu\nu}
\; .
\end{equation}
The mass matrix $\mathcal{M}$ depends only on the background
geometry and is constant in this case. As mentioned earlier, this matrix is uniquely determined
by unitarity arguments and requires tuning. 

There is a rather straightforward nonlinear completion of the leading infrared deformation \cite{ArkaniHamed:2002sp},
\begin{eqnarray}
\label{mdefg}
S_{_{\rm deform}}
&=&
-\frac{m^2}{2} \int_{M_1} {\rm d}^4 x \sqrt{-g_1}\; 
\mathcal{M}^{\alpha\beta\mu\nu}(g_1) \; H_{\alpha\beta} H_{\mu\nu} (g_1, g_2)
\; , \nonumber \\ \nonumber \\
H_{\alpha\beta}
&=&
g_{1\; \alpha\beta} - E^{\; \mu}_{\alpha} E^{\; \nu}_{\beta} \; g_{2\; \mu\nu}
\; ,  
\end{eqnarray}
where $E$ denotes the pullback from $M_2$ to $M_1$. Since the spacetimes
$(M_1,g_1)$ and $(M_2,g_2)$ need not be diffeomorphic to each other (even not
in the perturbative sense), $H(x)$ is in general not a fluctuation on $M_1$. 
In fact, the deformation (\ref{mdefg}) represents a mass term for a 
graviton\footnote{In abuse of notation and logic, we refer to metric perturbations
as gravitons irrespective of the background geometry.} 
only if $(M_2,g_2)$ is a copy of $(M_1,g_1)$ at the background level.

Consider the case when the most relevant deformation is a mass term.
Let us assume for simplicity, that some sort of Higgs mechanism 
is responsible for the graviton mass generation. For instance, in the Fierz--Pauli 
setup, a massless graviton, which has two transverse polarisation states, combines 
with a Goldstone vector to a massive spin-2 field, which has six polarisation
states in general (but only five on Minkowski spacetime). 
The Goldstone vector carries three transverse polarisation states
and one longitudinal polarisation state. 
When the massive graviton is at rest, its six polarisation states
are completely equivalent. However, if it is moving, the longitudinal polarisation 
becomes increasingly parallel to the graviton's momentum. As a consequence, 
at high energies, a massive graviton might look like the 
longitudinal polarisation state carried by the Goldstone vector \cite{ArkaniHamed:2002sp}. 
This statement is known as the Goldstone equivalence theorem and rests on the underlying
gauge invariance. 

The longitudinal polarisation state does not receive kinetic support from the
Einstein--Hilbert term, which is precisely why it is in the focus of unitarity 
requirements. As a matter of fact, many distinguished features of massive gravity,
like, for example the van Dam--Veltman--Zakharov discontinuity \cite{vanDam:1970vg} \cite{Zakharov:1970cc} , the 
Vainshtein radius \cite{Vainshtein:1972sx}, or the structure of the Fierz--Pauli mass term on Minkowski
spacetime \cite{Fierz:1939ix}, are captured by the longitudinal polarisation state of the Goldstone vector.  
Consistency invesitgations of massive gravity mostly focused on ghostly excitations 
at some finite perturbation level based on a Minkowski ground state, with the earliest
exception being Higuchi's unitarity analysis on de Sitter spacetime \cite{Higuchi:1986py}.
 
Higuchi found a consistency relation between the deformation parameter, the graviton mass,
and the curvature scale of de Sitter, set by the cosmological constant: In order to avoid negative norm states $m^2>H^2$, where $H$ denotes the Hubble constant. 
This bound is of great interest, since de Sitter geometry is unique in the sense
that is does not require any source specification. 
However, from a field theoretical point of view this makes the setup special, because
the background reference scale is constant here.

In this paper, we generalise Higuchi's bound from de Sitter to general Friedman 
cosmologies by employing the Goldstone equivalence theorem outlined above.
We find a competition between classical stability, the requirement that perturbations
respect the background, and quantum stability, the requirement that the spectrum does not 
contain negative norm states.   
For the special case of a de Sitter background, both criteria coincide and give rise to 
the unitarity bound quoted above. The situation is richer for generic Friedman cosmologies. 
There, the very nature of either bound is more intriguing, since it involves a time dependent 
curvature scale that is monotonously increasing or decreasing in the past, depending
on the specific sources that drive the background expansion. 
Generically, none of the bounds can be satisfied on the entire spacetime manifold. 
This, however, does not imply that the theory is invalidated. Indeed, it turns out that 
massive gravitons in generic Friedman universes are protected against unitarity violations.
More precisely, it is not clear at all whether a unitarity bound really exists, because before
entering the would-be unitarity violating spacetime region, the theory becomes strongly 
coupled \footnote{The connection between stability and strong coupling is clarified at the end of section \ref{sec_goldstone}. }. 

We reach the following verdict: Phenomenological constraints require to choose the
initial hypersurface close to the present hypersurface. 
In the case of a radiation or matter dominated Friedman universe, the evolution 
towards future 
hypersurfaces is guaranteed to be healthy (for consistent initial conditions)
by the strictly monotonic background expansion. 
In the most interesting case of a matter-cosmological constant mixture, the future evolution will be sound provided 
the mass is large enough and
consistent initial condition have been imposed. 
In all of the above cases, 
evolving backwards
in time, massive cosmological gravitons will soon enter a strong coupling regime that demands
a nonlinear completion of the theory. In other words, 
the fact that we have a sound theory on all past hypersurfaces is nontrivial and ensured 
by a strong coupling phenomenon that is confined to these spacetime regions.  
This is the advertised self-protection mechanism. 

Last but not least, we show that all conclusions hold for a generic
Friedman source.

%%%%%%%%%%%%%%%%%%%%%%%%%%%%%%%%

\section{Goldstone--St\"uckelberg analysis} \label{sec_goldstone}
In this section we derive the classical and quantum stability requirements for massive cosmological gravitons.

Consider two copies of a generic (background) spacetime, $(M_{\rm B}, \gamma)$ and $(M_{\rm B}, g)$ with 
$\gamma \equiv g + h$, where $h$  denotes the metric perturbation obeying $|g(t)| \gg |h(t,\bf{x})|$. 
In this case, $H=h$ is a perturbation (where, for simplicity, we have chosen the same coordinate system
on both manifolds) under the spell of diff$(M_{\rm B})$ for vanishing deformation 
parameter. 

Turning on the most relevant deformation (\ref{mdefg}),
the gauge symmetry is deformed to the diagonal subgroup of diff$(M_{\rm B}) \; \otimes$ diff$(M_{\rm B})$.
In other words, the deformation removes the freedom to gauge $h$ relative to the background
geometry.  The massive graviton, however, still carries six degrees of freedom, due to the second Bianchi identity
\begin{equation}
\label{const}
\nabla^\mu h_{\mu\nu} 
=
\nabla_\nu h
\; .
\end{equation}
In general, this is not a gauge choice. For instance, in the undeformed theory for massless gravitons
on Minkowski spacetime
the constraint (\ref{const}) is not a legitimate gauge, because the corresponding 
gauge shifts become singular for this choice (as a testimony of the van Dam--Veltman--Zakharov 
discontinuity on this background).

As is well known, in view of this explicit symmetry deformation, there are two equivalent state descriptions.
In the first case, the metric perturbation is split according to
\begin{equation}
h_{\mu\nu}
=
h^{\perp}_{\mu\nu} + \nabla_{(\mu} V_{\nu)}
\; ,
\end{equation}
where $h^{\perp}$ is covariantly conserved and carries two transverse degrees of freedom, while $V$ is unconstrained and carries four degrees of freedom. The latter can be decomposed further, $V=V^\perp + \partial \Psi$. Here $\nabla\cdot V^\perp= 0$ and $\Psi$
carries one degree of freedom. Using this state description, the theory can be considered
as a gauge fixed theory. 

The second and equivalent state description, called the Goldstone--St\"uckelberg completion, is based on adding four degrees of freedom, carried by a vector field $\pi$ in order 
to restore the original gauge symmetry, diff$(M_{\rm B}) \; \otimes$ diff$(M_{\rm B})$. 
In this case, the completion is given by
\begin{equation}
H_{\mu\nu} \equiv
h_{\mu\nu} + \nabla_{(\mu} \pi_{\nu)},	
\end{equation}	
where $H$ has ten degrees of freedom, of which six are carried by $h$ and four by 
the Goldstone--St\"uckelberg vector $\pi$, which can be further decomposed as
$\pi=\pi^\perp + \partial \phi$.
Here $\nabla\cdot\pi^\perp=0$ and $\phi$ carries one scalar degree of freedom. 

As mentioned above, the crucial point of this construction is the restored gauge symmetry
that allows to shift $h$ and $\pi$ relative to the background such that the 
Goldstone--St\"uckelberg completed $H$ itself is rendered gauge invariant. 
It is clear that four degrees of freedom represent gauge redundancies, leaving us with six
physical degrees of freedom. 

We choose to work with the Goldstone--St\"uckelberg 
completed state, for which the leading relevant deformation is exactly the celebrated
Fierz--Pauli mass term,
\begin{eqnarray}
\label{mass}
S_{_{\rm mass}}
&=&
-\frac{m^2}{2} \int_{M_B} {\rm d}^4 x \sqrt{-g}\; 
 H_{\alpha\beta}\; \mathcal{M}^{\alpha\beta\mu\nu}(g) \; H_{\mu\nu}
\; , \nonumber \\ \nonumber \\
&&\mathcal{M}_{\alpha\beta\mu\nu}(g)
=
 \; g_{\alpha\mu} g_{\beta\nu} -  \; g_{\alpha\beta}g_{\mu\nu}
\; .
\end{eqnarray}
The mode of the metric fluctuation corresponding to the
Goldstone--St\"uckelberg scalar $\phi$ dominates scattering processes at high momenta. 
This is tantamount to the Goldstone boson equivalence theorem and most easily understood
from the observation that this mode enters processes with at least two derivatives, 
$\nabla_\mu\partial_\nu\phi$, which, therefore, grows fastest in the high momentum limit. 
This is precisely the regime for which we are interested in studying the stability of 
the deformed theory.

The field $\phi$ mixes with the metric perturbation $h$ through the mass term (\ref{mass}),
\begin{equation}
2 m^2 \left( \Box h - \nabla^{\mu} \nabla^{\nu} h_{\mu \nu} \right) \phi 
= 
2 m^2 \left( R^{(0)}_{\mu \nu} (g) h^{\mu \nu} - R^{(1)}(g,h) \right) \phi \; ,
\end{equation}
where we have integrated by parts and introduced the Ricci tensor $R^{(0)}\equiv R$ 
evaluated on the background configuration $g$, as well as the Ricci scalar $R^{(1)}$ 
expanded to first order in $h$.

In order to eliminate the kinetic mixing term $R^{(1)}(g,h) \phi$, we carry out a conformal transformation
\begin{equation} 
\label{conformal_trafo}
\hat{g}_{\mu \nu} = \Omega^{2} g_{\mu \nu} \equiv (1+\omega)^{2} g_{\mu \nu},
\end{equation}
which, at the linear level, is equivalent to $\hat{h}_{\mu \nu} = h_{\mu \nu} + 2 \omega \, \eta_{\mu \nu} $. 
The Einstein--Hilbert term transforms as 
\begin{equation} 
\label{EH_conformal}
\sqrt{-g} \Omega^2 R 
= 
\sqrt{-\hat{g}} \left( \hat{R} - 6 \Omega^{-2} \; \hat{g}^{ab}  \partial_a \Omega \partial_b \Omega  \right) 
\; .
\end{equation}
In order to eliminate the mixing between $h_{\mu \nu} $ and $\phi $ we must choose $\Omega^2 = 1 - 2 m^2 \phi $ or, equivalently (since $\phi $ is a first order quantity in the expansion of $H_{\mu \nu}$), 
$\omega =  - m^2 \phi$. 

The conformal transformation (\ref{EH_conformal}) contributes a standard kinetic term for
the Goldstone--St\"uckelberg scalar, while the massive deformation (\ref{mass}) gives
rise to a non-standard kinetic contribution with the metric field replaced by the 
background Ricci tensor,
\begin{equation} 
\label{kinetic_phi_curvature}
2 m^2 \left( \left( \Box \phi \right)^2 - \nabla_{\mu} \nabla_{\nu} \phi \nabla^{\mu} \nabla^{\nu} \phi \right) 
= 
2 m^2 R_{\mu \nu} \partial^{\mu} \phi \partial^{\nu} \phi \; .
\end{equation}
The action for $\phi$ is given by
\begin{equation}
\label{actionphi}
S 
= 
\int {\rm d} ^4x \sqrt{-\hat{g}} \; 
\left( A \dot{\phi}^2 + B^{ij}(\partial_i\phi)(\partial_j\phi)+ \dot{\phi}D^i\partial_i\phi\right)
\; ,
\end{equation}
with
$A=2m^2(-3m^2g^{00}+R^{00})$, 
$B^{ij}=2m^2(-3m^2g^{ij}+R^{ij})$,
and 
$D^i=4m^2(-3m^2g^{i0}+R^{i0})$.
In general, these coefficients are spacetime dependent functions. 
Note that in (\ref{actionphi}) we have not displayed any potential terms (self-couplings) such as
$R(g) h \phi$, since quantum stability refers to the free evolution, and classical stability
relies on the kinetic terms at high momenta. 
The corresponding Hamilton density reads
\begin{equation}
\mathcal{H}
=
\frac{\pi^2}{4\sqrt{-\hat{g}} A} - \sqrt{-\hat{g}}B^{ij}(\partial_i\phi)(\partial_j\phi) \; ,
\end{equation}
where $\pi\equiv \delta\mathcal{L}/\dot{\phi}$. 
The Hamiltonian is unbounded from below for $A<0$ or $B^{ij}$ positive definite. 
 
In the case of a generic Friedman background geometry the action for the Goldstone--St\"uckelberg scalar reduces to 
\begin{equation}
\label{pactionfrw}
S 
= 
\int {\rm d}^4x \sqrt{-\hat{g}}\;  
\left( A(t) \dot{\phi}^2 + B (t) \left( \vec{\nabla} \phi/a \right)^2 \right) \;,
\end{equation}
where $A(t)=6m^2(m^2-\dot{H}-H^2)$, $B(t)=- 2 m^2 (  3 m^2 -  \dot{H} - 3 H^2 )$,
$H=H(t)$ denotes the Hubble parameter, and $a=a(t)$ the scale factor.  
These coefficients depend on the energy-momentum source curving the Friedman
background and, in particular, can change signs during the background evolution.
Classical stability requires $A>0$ and $B<0$. 

In order to study quantum stability of the Goldstone--St\"uckelberg scalar on 
a generic Friedman background, we need to transform (\ref{pactionfrw}) into normal form.
This requires the following 
field redefinition: $\phi\rightarrow f \phi$ with $\dot{f} = - C f/(2A)$, where $C\equiv \dot{A}-3 H B$.
The Lagrangian transforms as
\begin{equation}
f^{-2}\; \mathcal{L} 
= 
A \;  \dot{\phi}^2 - C \; \phi \dot{\phi} +  B \; \left( \vec{\nabla} \phi/a \right)^2 + C^2/(4A) \;  \phi^2  \; .
\end{equation}

The coefficient $A$ controls the sign between $\dot{\phi}$ and $f^{-2} \pi=2 A \dot{\phi}-C\phi$
and has therefore an important impact on the quantum stability of (\ref{pactionfrw}).
This can be worked out along the canonical quantisation prescription. 
As usual, we
postulate the equal time commutation relations
\begin{equation} 
\label{canonical_commutation}
\left[\phi(t, {\bf x}), \pi(t^\prime,\bf{x^\prime})\right]_{{t=t^\prime}}
=
i \delta^{(3)}\left(\bf{x}-\bf{x^\prime}\right) \;, \dots \;,
\end{equation}
and decompose the field $\phi(x)$ into modes 
$U(t,{\bf k})\equiv u(t,{\bf k})\exp{({\rm i}\bf{k}\cdot\bf{x})}$,
where $u(t,{\bf k})$ satisfies  
\begin{equation} 
\label{eom_phi_normal}
A \ddot{u} - \left[B\left(\frac{\bf k}{a}\right)^2 
+ 
A \left( \frac{{\rm d}}{{\rm d}t} + \frac{C}{2A} \right)\frac{C}{2A}  \right] u
= 0 
\; .
\end{equation}
The fact that the collection of $U(t,{\bf k})$ represents a complete 
orthonormal set of solutions (with respect to a spatial hypersurface-independent scalar product)
results in a simple condition on the Wronskian of the solutions,
\begin{equation} 
\label{phi_norm}
\left(u^* \dot{u} - u \dot{u}^*\right)(t,{\bf k}) = 1 \;.
\end{equation}
As a consequence, the Goldstone--St\"uckelberg scalar may be expanded as
\begin{equation}
\phi (t,{\bf x}) 
= 
\int \frac{{\rm d}^3 k}{(2\pi)^{3/2}} \frac{1}{\sqrt{2 |A|}  f} \;
\left(  u(t,{\bf k}) \exp{\left({\rm i}{\bf k}\cdot{\bf x}\right)} a({\bf k})
+  
 u^*(t,{\bf k})\exp{\left(-{\rm i}{\bf k}\cdot{\bf x}\right)}a^{\dagger}({\bf k})\right) \;.
\end{equation}
Inserting this expansion and the corresponding one for $\pi$ into the canonical
commutation relations (\ref{canonical_commutation}) yields
\begin{equation}
\left[ a({\bf k}), a^{\dagger}({\bf k')} \right] 
= 
\operatorname{sign}(A) \; \delta^{(3)}({\bf k}-{\bf k'}).
\end{equation}
The construction of a vacuum state and Fock space can now proceed as usual.
However, whenever $A<0$, the construction results in negative norm states,
which violate unitarity.  
 
For an arbitrary spacetime, this quantisation procedure bears conceptual challenges,
because there may be no Killing vectors at all to define positive frequency modes. 
The situation is simpler for a Friedman background since it accommodates a restricted set
of isometries, i.e. invariance under spatial rotations. Then, together with the corresponding 
Killing vectors there exist associated (natural) coordinates. 
Of course, coordinate systems are physically irrelevant --- a fact that renders the particle
concept somewhat arbitrary on curved spacetimes. However, this concerns the interpretation
of the theory. The unitarity requirement $A<0$ is a coordinate independent statement,
since Bogolubov transformations are unitary transformations. 

The most important result of this section is that stability considerations led us 
to require $A>0$ and $B<0$ in order to have a bounded Hamiltonian and, in addition,
$A>0$ to have a sound probabilistic interpretation. Hence, provided the theory respects
unitarity, $B<0$ represents the classical stability bound. 

This is also clear from the equation of motion (\ref{eom_phi_normal}). Indeed, for $A>0$,
to have a stable solution, requires $B<0$. Further, in order to have a damped solution
at late (early) times demands $C>0 (C<0)$. Whenever this condition is violated, the 
background will be destabilised. 

In order to investigate the signs of the coefficients $A,B,C$ in a spatially flat Friedman Universe,
consider the Friedman equations
\begin{eqnarray}
3H^2&=&8\pi G\rho+\Lambda ,\\
3\left(\dot{H}+H^2\right)&=&-4\pi G\left(\rho+3p\right)+\Lambda,
\end{eqnarray}
where $\rho$ is the density, $p$ the pressure, and $\Lambda$ the cosmological constant. An expanding universe is characterised by $H>0$
and $\dot{\rho} <0$, so $\dot{H}<0$. Furthermore, $\ddot{H} = -3 H\dot{H}(1+c_{\rm s}^{\; \; 2})$,  where $c_{\rm s}$ denotes the 
isentropic sound speed. As a consequence, $\ddot{H}>0$, for an equation of state governing an arbitrary mixture of matter and radiation.

In the absence of a cosmological constant, $H$ and all its time derivatives vanish at late times. The late time asymptotics of the coefficients 
are therefore given solely by the gravitons mass, $A=6m^4=-B>0$ and $C=0$. Consequently, the modes (\ref{eom_phi_normal}) are well-behaved. 
In a de Sitter universe, $A=6m^2(m^2-H^2)=-B$ and $C\propto A$. The absence of negative norm states requires $m^2>H^2$,
which is nothing else but the famous Higuchi bound. We see that in this case, classical and quantum stability collapses to a single criterion
between the graviton mass and the constant de Sitter curvature scale. 

For a generic Friedman universe, at any moment in time, $A>-B$ and 
$C=6m^2 H(3m^2 - \ddot{H}/H-3\dot{H}-3H^2) < 3HA$. 
As a consequence, evolving the modes 
(\ref{eom_phi_normal}) backward in time, $B$ and $C$ change signs before $A$ does (i.e. at later cosmological times). 
In the case of $C$ this sign change stabilises the modes since it leads to mode damping. This stabilisation, however, is only marginal
and nullified by the change of sign of $B$ which triggers an exponential instability that dominates the large proper momentum regime. 

As a consequence, evolving the modes backwards in cosmological times, the system enters first a strong coupling regime
(at later cosmological time) before it would violate unitarity (at earlier cosmological times). In this sense, the strong coupling
regime protects massive cosmological gravitons from unitarity violations.  In other words,  whenever massive gravitons in an expanding
Friedman universe experience unitarity violations, then, for sure, they are already in a strong coupling regime that demands a nonlinear
completion, but not vice versa. In this respect, de Sitter is a borderline geometry since both inconsistencies coincide.

It is important to appreciate that once a generic mode enters the classical instability region $B>0$, it does not destabilise the
background instantaneously. Instead, the characteristic time scale for this to happen is $T(k)\propto a/k$. 
As a consequence, modes with arbitrary large proper momentum become strongly coupled without further delay 
once $B>0$. More precisely, for arbitrary initial conditions, there will always be a critical proper momentum 
$k_{*}$ such that all modes with momenta $k>k_*$ enter the nonlinear regime before they would violate the unitarity bound.

Strictly speaking, the self-protection mechanism is confined to and efficient for $B\le 0$, 
that is, before the fluctuations enter the unstable regime, which is characterised by the background coefficient $B$ becoming positive, they hit a strong coupling regime. 
Indeed, for $B<0$ and $B\rightarrow 0$, the proper kinetic energy density $\propto B (\nabla \phi/a)^2$ becomes subdominant on any scale as compared to the potential energy. 
Hence, the exponential instability discussed above, characterized by $B>0$, 
is a testimony of this strong coupling regime and can legitimately be used to identify it.

%%%%%%%%%%%%%%%%%%%%%%%%%%%%%%%%%%%%%%%%

\subsection{Self-protection under matter impact}

In the following we argue that both the stability and unitarity behaviour of the theory are generically not altered by the specific choice of a matter action $S_{\rm matter} \bigl[\Psi_i, g_{\mu \nu} \bigr]$, where $\Psi_i$ denotes the collection of matter degrees of freedom.  
This is remarkable because the matter action explicitly depends on the metric field and, in particular, on the Goldstone--St\"uckelberg scalar
$\phi$. This coupling could, in principle, change the dynamics of $\phi$ in a relevant way such that the self-protection mechanism
will be overridden. As we will show below, this is not the case. 

Since the matter action is invariant under general coordinate transformations, the Goldstone--St\"uckelberg scalar enters the matter
sector only via the conformal transformation  (\ref{conformal_trafo}). Let us first expand the matter action
to second order in the fluctuations,
\begin{equation} 
S_{\rm matter} \left[ \Psi_i + \delta \Psi_i, g + h \right] 
\supset 
\int {\rm d}^4x \;  {\rm d}^4y \; h_{\mu \nu}(x) \; 
\left(h_{\alpha \beta}(y) \frac{\delta}{\delta g_{\alpha \beta}(y)} + \delta \Psi_i(y) \frac{\delta}{\delta \Psi_i(y)}\right) T^{\mu \nu}(x) \; .
\label{SM_second}
\end{equation}
After the conformal transformation (\ref{conformal_trafo}), $h=\hat{h} + 2 m^2 \phi g$,
the first term on the right-hand side of (\ref{SM_second}) contributes only
self-interaction terms of the form $\phi^2$ and couplings $h\phi$.
The coefficient of this potential term can be estimated to be of order $H^2$.
Hence, the potential term is subdominant as compared to the kinetic term 
at high proper momenta. 

The conformal transformation of the second term on the right-hand side of
(\ref{SM_second}) gives rise to a coupling between the Goldstone--St\"uckelberg scalar
and the matter degrees of freedom,
\begin{equation} 
\label{phi_psi_partial}
- 2m^2\int {\rm d}^4x \; \phi \frac{\partial T}{\partial (\partial_{\mu} \Psi_i)} \partial_{\mu} \delta \Psi_i \; ,
\end{equation}
where we have again neglected potential terms of the form $\phi \delta \Psi_i$.
A coupling like (\ref{phi_psi_partial}) will not modify the momentum field conjugated to $\phi$, and, hence,
the unitarity bound is robust against its inclusion. Suppose now we integrate (\ref{phi_psi_partial}) by 
parts, thereby producing a time derivative acting on $\phi$. This will modify the conjugated momentum
field, but it will only contribute a term proportional to $\delta \Psi_i$. Again, such a coupling respects 
the unitarity bound.

With respect to stability, the coupling (\ref{SM_second}) might alter the dynamics of the 
Goldstone--St\"uckelberg field substantially, as it represents a derivative coupling to the matter degrees
of freedom. However, the coefficient of this derivative coupling is a vector field constructed solely from
background quantities. Due to the isometries of the Friedman geometry, this background vector 
has to be of the form
\begin{equation}
\left( \frac{\partial T}{\partial (\partial_{\mu} \Psi_i)} \right) \propto  e_{\mu}\equiv(1,0,0,0)
\; .
\end{equation}
Therefore, no spatial derivative enters the coupling (\ref{phi_psi_partial}). Thus the modes
experience an additional source proportional to $\partial_t \delta \Psi_i$, and as long as
the matter sector takes good care of itself this source cannot alter the stability bound, in particular
at large proper momenta.

%%%%%%%%%%%%%%%%%%%%%%%%%%

\section{Discussion}
Let us consider the concrete expressions for the coefficients $A$ and $B$ and discuss the possible implications of the corresponding bounds.

\subsection{De Sitter spacetime}
For a de Sitter spacetime, $A= 6m^2 (m^2 - H^2)=-B$. The absence of negative norm states requires the unitarity 
bound $m^2>H^2$ to hold, which is just the well-known Higuchi bound (in our conventions). 
Provided this bound has been satisfied, classical stability is established automatically, as $A=-B$. In fact, de Sitter geometry
is unique within the class of Friedman spacetimes where classical and quantum stability requirements coincide, and
where stability is solely expressed in terms of two model parameters, i.e. the graviton mass and the constant
de Sitter curvature scale. Since linearisation is permissible, the stability bound, $m^2>H^2$, should be an important
consistency condition for a nonlinear completion, as well.   
 
\subsection{Generic Friedman spacetimes}
For a generic Friedman spacetime, the absence of negative norm states demands
$m^2 > H^2 + \dot{H}$, while classical stability requires $m^2  > H^2 + \dot{H}/3$.
We were able to confirm these stability bounds by a full-fledged perturbation analysis of massive cosmological
gravitons involving all degrees of freedom, as well as a complete set of couplings. (See \cite{cosmological_perturbation}.) The same bounds were also derived for the special case of scalar field matter in \cite{Grisa:2009yy}. The findings of  \cite{Blas:2009my}, however, do not coincide with ours, due to the unconventional matter Lagrangian that has been used in their paper.

In the case of an expanding Universe the unitarity bound will always
be satisfied during radiation ($m^2>-H^2$) or matter domination ($m^2>-H^2/2$). Thus, the absence of negative norm states
is guaranteed by the isotropic expansion. This does, however, not imply that classical stability is unchallenged during
the Universe's expansion, since the classical stability bounds $m^2>H^2/3$ (during radiation domination) and $m^2>H^2/2$
(during matter domination) will eventually be violated when the modes are being evolved backwards in cosmological time. 

The situation becomes more interesting for a universe filled with a mixture of matter and a cosmological constant. 
In this case, there is no unitarity issue at early times before the transition from matter to cosmological constant domination.
However, after the matter-cosmological constant transition $H^2+\dot{H}$ will become positive and eventually constant
as the Universe evolves towards its de Sitter fate. If the cosmological graviton's mass is larger than the asymptotic 
value for the Hubble parameter set by the de Sitter curvature scale, the theory always respects unitarity. 
If, however, the mass parameter is chosen to be smaller than the cosmological constant, it is guaranteed that the $\phi $ modes first enter the epoch of classical instability, before (evolving forward in time) they hit the then would-be unitarity bound (since $\dot H < 0 $). 
As a consequence, massive cosmological gravitons are protected against unitarity violation by the background expansion. 
We conjecture that this sort of self-protection could be robust against any nonlinear completion.

We are well aware that a non-linear completion may suffer from another inconsistency problem, the so-called Boulware--Deser ghost \cite{Boulware:1973my}. 
This definitely requires further investigations, but there are already first indications that certain 
non-linear deformations could be consistent \cite{deRham:2010ik,Gabadadze:2009ja}.   

Similar remarks apply to more general source terms. For an expanding universe, $\dot{H}<0$, so that the massive 
cosmological gravitons always first enter the strong coupling regime, rendering the would-be unitarity bound
fictitious and opening an exciting window of opportunity towards a self-protection mechanism. 
Finally, let us state the situation is reversed in a contracting universe, where $\dot{H}>0$. 
Here, negative norm $\phi$ states will show up in the weak coupling regime. There is no obvious self-protection
mechanism in this case.

%%%%%%%%%%%%%%%%%%%%%%%%%%%%%%%%%%%%%%%%%%%%%%%%%%%%%%%%%%%%%%%%%%%%%%%%%%

\section*{Acknowledgments}

The authors would like to thank Eugeny Babichev, Cliff Burgess, Stanley Deser, Gia Dvali, Justin Khoury, Michael Kopp, Florian Niedermann, and Andrew Tolley for helpful and inspiring discussions.
DDD acknowledges gratefully the hospitality of the Arnold Sommerfeld Center and the Excellence Cluster Universe.
The work of DDD was supported by the Danish Natural Science Research Council.
The work of SH was supported by the DFG cluster of excellence 'Origin and Structure of the Universe'
and by TRR 33 'The Dark Universe'.

%%%%%%%%%%%%%%%%%%%%%%%%%%%%%%%%%%%%%%%%%%%%%%%%%%%%%%%%%%%%%%%%%%%%%%%%%%

\end{document}